\documentclass[article,twocolumn,preprintnumbers,amsmath,amssymb,floatfix,superscriptaddress]{revtex4-1}
\usepackage{graphicx}
\usepackage{rotating}
\usepackage{comment}
\usepackage{dcolumn}
\usepackage{makeidx}
\usepackage{color}
\usepackage{amsmath}
\usepackage{amsfonts}
\usepackage{mathrsfs}
\usepackage{wrapfig}
\usepackage{braket}
\usepackage{bm}
\usepackage{float}
\usepackage{url}

\usepackage{dcolumn}
\usepackage{bm}
\usepackage{tabularx}
\usepackage{hyperref}
\usepackage{cleveref}

\usepackage{amsmath,amssymb,amsfonts,bbm,graphicx}
\usepackage[latin1]{inputenc}
\usepackage[T1]{fontenc}
\usepackage{verbatim}
\usepackage{textcomp}
\usepackage{graphicx}
\usepackage{amsmath}

\usepackage{graphicx}
\usepackage{dcolumn}
\usepackage{bm}
\usepackage{xcolor}

\usepackage{hyperref}
\hypersetup{colorlinks=true, linkcolor=blue, citecolor=blue, urlcolor=blue}

\usepackage[normalem]{ulem}

\begin{document}

\title{On the microscopic mechanisms behind hyperferroelectricity}

\author{Mohamed Khedidji}
\affiliation{Facult\'e de Chimie, Universit\'e Houari Boumedienne, BP.32, El Alia, Bab Ezzouar, Alger, Algeria}
\author{Danila Amoroso}
\affiliation{National Research Council CNR-SPIN, c/o Universit\`a degli Studi ``G. D'Annunzio'', I-66100 Chieti, Italy}
\author{Hania Djani}
\affiliation{Centre de D\'eveloppement des Technologies Avanc\'ees, Cit\'e 20 ao\^ut 1956, Baba Hassen, Alger, Algeria}


\begin{abstract}
Hyperferroelectrics are observing a growing interest thanks to their unique property to retain a spontaneous polarization even in presence of a depolarizing field, corresponding to zero macroscopic displacement field ($\mathscr{D}=0$) conditions. 
Hyperferroelectricity is ascribed to the softening of a polar $LO$ mode, but the microscopic mechanisms behind this softening are not totally resolved. 
Here, by means of phonon calculations and force constants analysis, performed in two class of hyperferroelectrics, the ABO$_3$-LiNbO3-type systems and the hexagonal-ABC systems, we unveiled the common features in the dynamical properties of a hyperferroelectric that are leading the $LO$ instability:  negative or vanishing on-site force constant associated to the cation driving the $LO$ polar mode and a destabilizing cation-anion interactions; both induced by short-range forces. We also predicted a possible enhancement of the hyperferroelectric properties under increasing external positive pressures: the pressure strengthens the destabilizing short-range interactions, inducing a stronger $LO$ mode instability and the increase of the longitudinal mode effective charges associated to the unstable $LO$ mode, suggesting an eventual enhancement of the  $\mathscr{D}=0$ polarization, under compressive strain. 

\end{abstract}
\maketitle

\section{Introduction}
The concept of \emph{hyperferroelecticity} (hyperFE) was first introduced in semi-conducting hexagonal ABC ferroelectrics (FE) by Garrity and coauthors in Ref.~\cite{Garrity2014}. 
The prefix \emph{hyper-} is related to the intrinsic property of such a new class of proper ferroelectrics to display a persistent polarization even in the presence of a depolarization field; something unachievable by standard ferroelectrics (FE)~\cite{Zhong1994,Garrity2014}.
In fact, by analyzing the electric equation of state, Garrity \textit{et al.} showed that, 
in contrast to standard ferroelectrics which spontaneously polarize only under zero macroscopic electric field ($\mathscr{E}=0$), hyperferroelectrics can spontaneously polarize under both zero macroscopic electric field ($\mathscr{E}=0$) and zero macroscopic displacement field ($\mathscr{D}=0$), i.e. unscreened depolarization field under open circuit boundary conditions. Such features thus make hyperFE systems suitable for applications as low dimensional functional materials~\cite{FH2014, PhysRevB.96.235415, karin2020, PhysRevLett.117.076401,halides}; moreover, the existence of a switchable electric polarization in hyperFEs can be allowed in metals and not restricted to only insulators and semiconductors as for standard FEs ~\cite{xiang2014,Pengfei2016,PhysRevB.96.235415}.


The difference in behavior between FE and hyperFE systems stems from the different type of lattice instabilities displayed in the paraelectric phase, as it was first pointed out by Garrity \textit{et al.}: the well-known FE instability is related to an unstable zone-center transverse optic ($TO$) mode~\cite{PhysRevB.42.6416, rabe2007}; the hyperFE instability is related to an unstable zone-center longitudinal optic ($LO$) mode beside the unstable $TO$ one. 
Moreover, they ascribed the appearance of such unstable $LO$ mode in the narrow-gap ABC hyperFEs, to a small $LO$-$TO$ splitting resulting from the small Born effective charges (BEC) and relatively large electronic contribution to the dielectric constant $\epsilon^{\infty}$~\cite{Bennett_ABC}. 
Nevertheless, in a later work, Li \textit{et al.}~\cite{Pengfei2016} reported unstable $LO$ modes also in some LiNbO$_3$-type ferroelectrics showing, on the contrary, anomalous BEC and small $\epsilon^{\infty}$. This apparent contradiction motivated these authors to provide a further insight into the microscopic mechanisms behind the $LO$ mode instability. By modeling the LiNbO$_3$-type systems via an effective Hamiltonian, they identified the structural instabilities related to Li atoms, driven by short-range interactions, as the origin of the hyperferroelectricity. Even So,  the identification of the common microscopic origin behind the softening of this $LO$ mode is still under debate. 

In this work, by means of Density Functional Theory (DFT) calculations and the  analysis of the dynamical properties obtained from Density Functional Perturbation Theory (DFPT), we confirm and discuss in detail the direct relationship between unstable $LO$ mode and destabilizing short-range (SR) interactions in  ABO$_3$-LiNbO$_3$-type oxides (with A=Li,Na and B=Ta,Nb,V), 
extending our findings  to the hexagonal ABC systems, LiBeSb, LiZnP, LiZnAs and NaMgP compounds, as representative examples. 
In particular, the exploration of the real-space on-site and interatomic force constants (IFCs) allowed us to distinguish between specific contributions of the long-range (LR) and sort-range (SR) forces to the interatomic interactions that are, in turn, related to the structural properties of the investigated systems. 
Our study reveals common microscopic mechanisms driving  hyperferroelectricity: a structural frustration, arising from the under-coordination of the small sized A atom in  LiNbO$_3$-type systems and small sized B atom in the hexagonal ABC, induces 
the off-centering of the frustrated cations towards the neighboring out-of-plane anions. Such polar distortion is not only driven by LR dipole-dipole interactions but also by SR interactions; the first one contributing to the ferroelectric ($TO$ mode) instability and the second to the hyperferroelectric ($LO$ mode) one. 
Additionally, we also investigated the effect of external isotropic pressure on the LiNbO$_3$-type systems, finding out a possible enhancement of hyperferroelectricity.

\begin{figure}[t!]
\includegraphics[width=8.2cm,keepaspectratio=true]{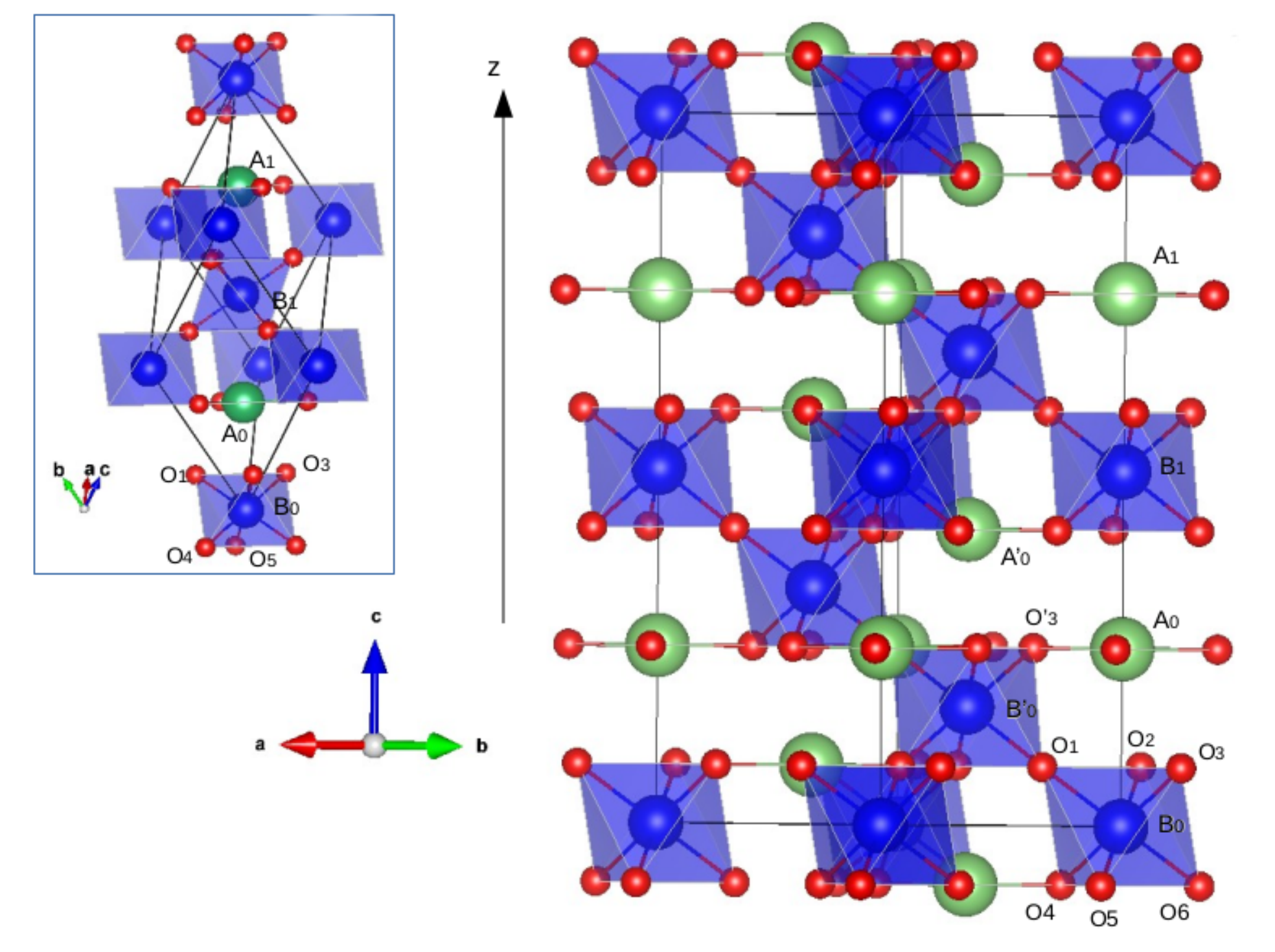}
\caption{ The paraelectric $R\bar{3}c$ structure of ABO$_3$-LiNbO$_3$-type oxides in its Conventional hexagonal unit cell (the primitive rhombohedral unit cell inset). A atoms are in Green color, B atoms in Blue and Oxygen atoms in red.}
 \label{structure_1}
\end{figure}

\section{Methods}
Calculations were performed within DFT ~\cite{PhysRev.136.B864, PhysRev.140.A1133} using a plane waves method as implemented in the ABINIT package ~\cite{abinit3, gonze09}. The exchange correlation energy functional was evaluated within the generalized gradient approximation (GGA) employing the revised Perdew-Burke-Ernzerhof functional PBEsol~\cite{Perdew2008, dojo}. 
The wave functions were expanded up to a kinetic energy cutoff of 45 Hartrees. Integrals over the Brillouin zone were approximated by sums on a a 6$\times$6$\times$6 Monkhorst-Pack $k$-points mesh~\cite{monkhorst76}. We relaxed the structure until the remaining forces on the atoms were less than $10^{-5}$ Hartree/Bohr and the stresses on the unit cell smaller than $10^{-7}$. Phonons frequencies, IFCs, Born effective charges and dielectric tensors were computed on the primitive paraelectric $R\bar{3}c$ phase using density-functional perturbation theory (DFPT) \cite{baroni01,gonze97}. Note that the high temperature paraelectric $R\bar{3}c$ phase is experimentally observed for LiTaO$_3$ and LiNbO$_3$ and hypothetical for LiVO$_3$ and NaVO$_3$, but considered here to analyze trends and understand mechanisms.
We also performed relaxation and DFPT calculations on hexagonal ABC systems in the paraelectric cubic $P6_3/mmc$ phase, using GGA-PBEsol functional, with a 6$\times$6$\times$6 Monkhorst-Pack mesh and a kinetic energy cutoff of 45 Hartrees.

\section{Results}

In order to provide a clear understanding of mechanisms at play
in HyperFEs, we proceeded step-by-step:
first, we presented the structural properties of LiTaO$_3$, LiNbO$_3$, LiVO$_3$ and NaVO$_3$ in the centrosymmetric paraelectric $R\bar{3}c$  phase and their relationship with the associated low-symmetry polar $R3c$ phase.
Then, we discussed the dynamical properties of the  paraelectric phase. In particular, we reported phonons modes at the $\Gamma$-point, with a special focus on phonons associated to polar displacements along the cartesian $z$-direction -corresponding to the $R\bar{3}c$ trigonal axis- that are  driving the observed ferroelectric phase transition in LiTaO$_3$ and LiNbO$_3$. Phonons frequencies, eigendisplacements and mode effective charges are reported, together with the real space on-site and interatomic force constants, That are of particular importance here to reveal the interatomic interactions behind the softening of the $LO$ mode.


\subsection{Structural properties}

\begin{figure}[t]
\centering
\includegraphics[width=6cm]{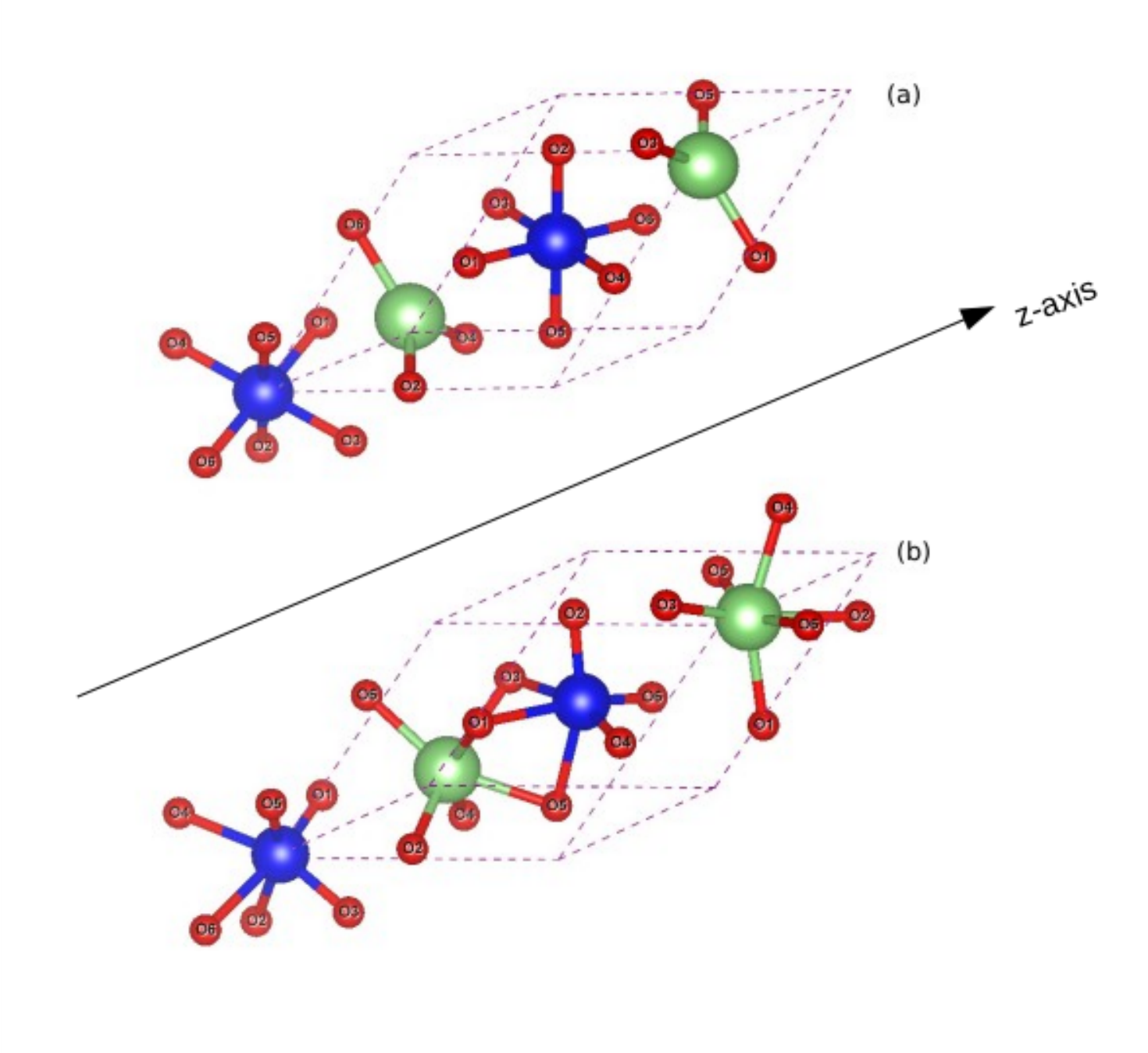}\\
\caption{(a) In the paraelectric $R\bar{3}c$ phase, Li is under-coordinated (surrounded with three in-plane oxygens). (b) In the ferroelectric $R3c$, the Li displace toward the out-of-plane oxygens cage along the $z$-direction: its coordination is then optimized from III to VI.}
\label{structure_1b}
\end{figure}

In ABO$_3$-LiNbO$_3$-type oxides, the paraelectric structure of $R\bar{3}c$ symmetry (Fig.~\ref{structure_1}) counts 10 atoms in its rhombohedral primitive cell (or 30 atoms in the hexagonal conventional cell) (See~\cite{supplemental}).  The atomic arrangement consists of chains of equidistant A-site (Li, Na) and B-sites (Ta, Nb, V) atoms along the trigonal axis (cartesian $z$-direction). As illustrated in Fig.~\ref{structure_1}, the transition-metal B atoms occupy the center of oxygen octahedra and the A atoms sit at the center of the in-plane nearest neighbors O-triangle and have six further next near neighbors out-of-plane oxygens (out-of-plane O$_{1,2,3}$, equivalent to O$_{4,5,6}$) (see also Fig.~\ref{structure_1b}(a)).
The ferroelectric structure of $R3c$ symmetry originates from  the off-centering of B and A atoms along the trigonal axis (see Fig.~\ref{structure_1b}(b)). In particular, the Li-O polar displacement tends to improve the Li coordination environment by coming closer to three  of the six out-of-plane oxygens (Li-O$_1$) and moving away from the three in-plane oxygens (Li-O'$_3$), as reported in  Table~\ref{tab:1}.


\begin{table} [htpb!]
\begin{center}
\caption{Distances of Li-O bonds (\AA) in the paraelectric $R\bar{3}c$ and ferroelectric $R3c$ phases. During the ferroelectric $R\bar{3}c$ to $R3c$ transition, A atoms move away from $O'_3$ (the $A_0$-$O'_3$ distances increase) and get closer to $O_1$ (the $A_0$-$O_1$ distances decrease). Obviously, Na atoms show a lower ability to move away from the in-plane $O'_3$ triangle.  }
\label{tab:1}
\begin{tabular}{lcccccccccccccccc}
\hline\hline
    & \multicolumn{2}{c}{LiTaO$_3$}& &\multicolumn{2}{c}{LiNbO$_3$}& &\multicolumn{2}{c}{LiVO$_3$}& &\multicolumn{2}{c}{NaVO$_3$} & \rule[-1ex]{0pt}{3.5ex}\\
 \cline{2-3}\cline{5-6}\cline{8-9}\cline{11-12}
Bond & Para & Ferro & & Para & Ferro && Para & Ferro && Para & Ferro \rule[-1ex]{0pt}{3.5ex}\\
\hline
$A_0$-$O'_3$ & 1.97 & 2.04  && 1.96 & 2.04 && 1.96 & 2.04 && 2.39 & 2.40 \rule[-1ex]{0pt}{3.5ex}\\
$A_0$-$O_1$ & 2.79 & 2.28 && 2.79 & 2.28 && 2.61 & 2.14 && 2.62 & 2.46 \\
\hline\hline
\end{tabular}
\end{center}
\end{table}


\begin{table}[htpb!]
\caption{Unstable modes at $\Gamma$ in the paraelectric $R\bar{3}c$ phase. $\Gamma_2^{-}$ is polar along the $z$-direction (trigonal axis),  $\Gamma_2^{+}$  is antipolar along $z$-direction, $\Gamma_3^{-}$  is polar and doubly degenerated along $xy$-direction and  $\Gamma_3^{+}$ is antipolar and doubly degenerated along the $xy$-direction. Our results are in agreements with previous works~\cite{veithen2002, parlinski, toyoura, Pengfei2016}.}
\label{tab:2}
\centering
\begin{tabular}{lcccccc}
\hline\hline
mode irreps &&  LiTaO$_3$ & LiNbO$_3$ & LiVO$_3$ & NaVO$_3$\\
\hline
$\Gamma_2^{-}$ (A$_{2u}$) && 164$i$  & 200$i$ & 448$i$, 137$i$ & 459$i$\\
$\Gamma_2^{+}$ (A$_{2g}$) &&  96$i$ & 102$i$ & 101$i$& -\\
$\Gamma_3^{-}$  (E$_{u}$)&&  -& 95$i$& 386$i$& 442$i$\\
$\Gamma_3^{+}$ (E$_{g}$)&&  - & - & -& 279$i$, 7$i$\\
\hline\hline
\end{tabular}
\end{table}

\begin{table} [htpb!]
\begin{center}
\caption{Born effective charges (BEC)  and $\varepsilon_\infty$. The nominal valence charges of A, B and O are $+$1, $+$5 and $-$2, respectively. Only the diagonal elements are reported, the complete Table is given in Supplemental Material (See~\cite{supplemental}).}
\label{tab:3}
\begin{tabular}{lcccccccccccccccccccccccc}
\hline\hline

    & & & & \multicolumn{5}{c}{LiTaO$_3$}& & &\multicolumn{5}{c}{LiNbO$_3$} & \rule[-1ex]{0pt}{3.5ex} \\
 \cline{5-9}\cline{12-16}
Atom & & & & $Z_{xx}^{*}$ && $Z_{yy}^{*}$ && $Z_{zz}^{*}$ & & & $Z_{xx}^{*}$ && $Z_{yy}^{*}$ && $Z_{zz}^{*}$  & \rule[-1ex]{0pt}{3.5ex}\\
\hline
A & & & & 1.14 && 1.14 && 1.11 & & & 1.15 && 1.15 && 1.10 & \rule[-1ex]{0pt}{3.5ex}\\ 
B & & & & 7.67 && 7.67 && 8.33 & & & 8.33 && 8.33 && 9.19 & \\
O$_1$ & & & & $-$2.34 && $-$3.85 && $-$3.15 & & & $-$2.47 && $-$3.87 && $-$3.43 & \\
O$_2$ & & & & $-$4.12 && $-$1.75 && $-$3.15 & & & $-$4.54 && $-$1.80 && $-$3.43 & \\
O$_3$ & & & & $-$2.34 && $-$3.53 && $-$3.15 & & & $-$2.50 && $-$3.85 && $-$3.43 & \\
\hline
  & & & & $\varepsilon_\infty^{xx}$ && $\varepsilon_\infty^{yy}$ && $\varepsilon_\infty^{zz}$ & & & $\varepsilon_\infty^{xx}$ && $\varepsilon_\infty^{yy}$ && $\varepsilon_\infty^{zz}$ & \rule[-1ex]{0pt}{3.5ex}\\
  \cline{5-9}\cline{12-16} \rule[-1ex]{0pt}{3.5ex}
$\varepsilon_\infty$ & & & & 5.20 && 5.20 && 5.63 & & & 6.12 && 6.12 && 6.80 & \rule[-1ex]{0pt}{3.0ex}\\
\hline
       & & & & \multicolumn{5}{c}{LiVO$_3$}& & & \multicolumn{5}{c}{NaVO$_3$} & \rule[-1ex]{0pt}{3.5ex} \\
 \cline{5-9}\cline{12-16}
Atom & & & & $Z_{xx}^{*}$ && $Z_{yy}^{*}$ && $Z_{zz}^{*}$ & & & $Z_{xx}^{*}$ && $Z_{yy}^{*}$ && $Z_{zz}^{*}$ & \rule[-1ex]{0pt}{3.5ex}\\
\hline
A & & & & 1.14 && 1.14 && 1.12 & & & 1.04 && 1.04 && 1.05 & \rule[-1ex]{0pt}{3.5ex}\\ 
B & & & & 11.20 && 11.20 && 12.36 & & & 13.13 && 13.13 && 12.80 & \\
O$_1$ & & & & $-$3.11 && $-$5.10 && $-$4.50 & & & $-$3.50 && $-$6.00 && $-$4.61 & \\
O$_2$ & & & & $-$6.10 && $-$2.11 && $-$4.50 & & & $-$7.20 && $-$2.30 && $-$4.61 & \\
O$_3$ & & & & $-$3.11 && $-$5.10 && $-$4.50 & & & $-$3.50 && $-$6.00 && $-$4.61 & \\
\hline
  & & & & $\varepsilon_\infty^{xx}$ && $\varepsilon_\infty^{yy}$ && $\varepsilon_\infty^{zz}$ & & & $\varepsilon_\infty^{xx}$ && $\varepsilon_\infty^{yy}$ && $\varepsilon_\infty^{zz}$ & \rule[-1ex]{0pt}{3.5ex}\\
  \cline{5-9}\cline{12-16} \rule[-1ex]{0pt}{3.5ex}
$\varepsilon_\infty
$ & & & & 13.10 && 13.10 && 14.40 & & & 15.61 && 15.61 && 14.53 & \rule[-1ex]{0pt}{3.0ex}\\
\hline\hline
\end{tabular}
\end{center}
\end{table}

\subsection{Phonon properties at $\Gamma$ and interatomic forces constants}


Within the harmonic approximation, structural instabilities are associated to negative curvature of the internal energy with respect to specific atomic displacements, yielding imaginary phonon frequencies~\cite{PhysRevB.36.6631, PhysRevB.49.5828, PhysRevB.52.6301}. In line with the \emph{``soft-mode theory''} first introduced by Cochran~\cite{PhysRevLett.3.412}, the ferroelectric transition is  ascribed to an unstable zone-center transverse optic ($TO$) phonon in the parent paraelectric phase associated to a polar atomic pattern of distortion; such a ``ferroelectric''  instability results from the delicate competition between stabilizing short-range (SR) forces and destabilizing long-range (LR) Coulomb interaction taking the form of a dipole-dipole (DD) interaction. 
In the following, we show that the ``hyperferroelectricity'' is rather resulting from an unstable zone-center longitudinal optic ($LO$) phonon driven by destabilizing SR forces. \\

Within the DPFT approach, the calculation of the interatomic force constants (IFCs), $C_{\alpha,\beta}(lk,l'k')$, and the analysis of the distinct SR and LR contributions, as defined in Refs~\cite{Ghosez_1996}, allow to identify which driving forces lead the system to exhibit eventual instabilities~\cite{PhysRevB.60.836,PhysRevB.97.174108}. In particular, within the used convention, the IFCs relates the $\alpha$-component of the force $F_{\alpha}(lk)$ on atom $k$ in cell $l$, to the induced displacement $\tau_{\beta}(l'k')$ of atom $k'$ in cell $l'$, through the expression 
$F_{\alpha}(lk)=-C_{\alpha,\beta}(lk,l'k')\tau_{\beta}(l'k')$~\cite{PhysRevB.60.836}:
if the induced force on atom $k$ is opposite to the direction of the displacement of atom $k'$, a discordant cooperative atomic motion takes place, eventually producing break of the spatial inversion symmetry and so the creation of a dipole moment;
accordingly, the IFC is positive and corresponds to a destabilizing interaction. Differently, the force on a single atom resulted from its isolated displacement from its initial crystalline position is specified by the ``on-site'' force constant; this can be written as a sum over IFCs: $C_{\alpha,\beta}(lk,lk)=-\sum'_{l'k'}C_{\alpha,\beta}(lk,l'k')$~\cite{PhysRevB.60.836}. In this case, a negative on-site force constant means an instability against isolated atomic displacement: the induced force and the atomic displacement are concordant, favoring thus the off-centering from the initial position; at opposite, a positive value means stability against isolated atomic displacements, as the induced force will bring the atom back to its initial position. 
Accordingly, we reported in what follow the phonon properties at $\Gamma$-point and the interatomic forces constants calculated in the primitive cell of the paraelectric $R\bar{3}c$ phase of LiTaO$_3$, LiNbO$_3$, LiVO$_3$ and NaVO$_3$. \\

\begin{table*}[htpb!]
\begin{center}
\squeezetable
\caption{Calculated phonon frequencies $\omega$ (cm$^{-1}$) of $\Gamma_2^{-}$ $TO$- and $LO1$- modes with their corresponding normalized eigendisplacements (in a.u.) and mode effective charges $\bar{Z}^{*}_m$. The Callen longitudinal mode effective charge  ($\bar{Z}^{*L}_{LO1}$) for $LO1$ is given between parenthesis.  Decomposition of the phonon frequency into the long-range (LR) and short-range (SR) contributions is also reported ($\omega^2$= $\omega_{LR}^2$ + $\omega_{SR}^2$).}
\label{tab:4}
\begin{tabular}{llccccccccccccccccccccccccc}
\hline
\hline
&modes & $\omega$ & A & B &  O$_{1/2/3}$ && $\bar{Z}^{*}_{m}$& $\omega^{2}$ &$\omega^{2}_{LR}$&$\omega^{2}_{SR}$\\  
\hline
LiTaO$_3$&$TO1$ &[164$i$] &+0.2081 &+0.0080 & $-$0.0605 && 4.64 & $-$27056 &+29634  & $-$56690\\ 
&$TO2$ &[149] &+0.1611 &$-$0.0239 & +0.0669 && -4.42 & +22203 &$-$171327 & +193531\\ 
&$TO3$ &[512]&+0.0254 &+0.0022 & $-$0.0120 && 6.47 & +262540  &$-$581901   & +844442 \\
&$LO1$ &[28$i$]&+0.2594 & $-$0.0105 &+0.0021 && (0.077) &- &-  & -\\
\hline
LiNbO$_3$&$TO1$ &[201$i$] &+0.1544 &+0.0268 & $-$0.0742 && 7.79 & $-$40285 &$-$187424  & +147138\\ 
&$TO2$ &[69] &+0.2093 &$-$0.0357 & +0.0389 && -3.60 & +4803 & $-$177323  & +182127\\ 
&$TO3$ &[466]&+0.0318 &$-$0.0011 & $-$0.0024 && 6.38 & +217662  &$-$473571  & +691233\\
&$LO1$ &[77$i$]&+0.2570& $-$0.0162 & $-$0.0057 && (0.052) &- &-  & -\\
\hline
LiVO$_3$&$TO1$ &[448$i$] &+0.0216 &+0.0651 & $-$0.0722 && 19.70 & $-$200861 &$-$980925  & +780064\\ 
&$TO2$ &[137$i$] &+0.2573 &$-$0.0218 & $-$0.0140 && 0.43 &  $-$18810 &  +56318  & $-$75128\\ 
&$TO3$ &[498]&+0.0262 &$-$0.0194 & $+$0.0168 && 4.71 & +248701  &$-$158808  & +407510\\
&$LO1$ &[138$i$]&+0.2536 &$-$0.0118& $-$0.0240 && (-0.008) & - &-  &  -\\
\hline
NaVO$_3$&$TO1$ &[459$i$] &+0.0021 &+0.0665 & $-$0.0716 && 20.18 & $-$210723 &$-$1161238& +950515\\ 
&$TO2$ &[159] &+0.1328 &$-$0.0315 & $-$0.0301 && 1.20 & +25365 & +14090 & +11274\\ 
&$TO3$ &[538]&+0.0033 &$-$0.0173 & +0.0167 && 4.67 & +290176  &$-$190507  &+480683 \\
&$LO1$ &[154]&+0.1318 &$-$0.0390&$-$0.0216 && (-0.034) & - &-  & -\\
\hline
\hline
\end{tabular}
\end{center}
\end{table*}

\begin{table} [htpb!]
\begin{center}
\caption{Overlap matrix elements $\langle \eta^{LO}|M|\eta^{TO}\rangle$($M=M_{\kappa}\delta_{\kappa \kappa'}$) between $TO$ modes eigenvectors and $LO1$ mode in the $R\bar{3}c$ paraelectric phase. 
}
\label{tab:5}
\begin{tabular}{lccccccccccccccccccccccc}
\hline
\hline
        & & & \multicolumn{2}{c}{LiTaO$_3$}&&\multicolumn{2}{c}{LiNbO$_3$}&&\multicolumn{2}{c}{LiVO$_3$}&&\multicolumn{2}{c}{NaVO$_3$} \rule[-1ex]{0pt}{3.5ex} \\
\cline{4-5}\cline{7-8}\cline{10-11}\cline{13-14}
& & &  & $LO1$ & &  & $LO1$ & &&  $LO1$ &&& $LO1$ &&\\
\hline
$TO1$& & &  & 0.73 & &&  0.52 & && 0.06 &&&0.11 && \\
$TO2$& & &  & 0.68 & &&  0.85 & && 0.99&&&0.99&&\\
$TO3$ & & &  & 0.08 & &&  0.06& && 0.01&&&0.03&&\\
\hline
\hline
\end{tabular}
\end{center}
\end{table}

\begin{table*}[t]
\centering
\caption{On-site force constants in (Hartree/bohr$^{2}$) related to different atoms calculated in the $R\bar{3}c$ paraelectric phase.}
\label{tab:6}
\begin{tabular}{lccccccccccccccccccccccc}
\hline
\hline
        & & & \multicolumn{3}{c}{\scriptsize LiTaO$_3$}&&\multicolumn{3}{c}{\scriptsize LiNbO$_3$}&&\multicolumn{3}{c}{\scriptsize LiVO$_3$}&&\multicolumn{3}{c}{\scriptsize NaVO$_3$} \rule[-1ex]{0pt}{3.5ex} \\
\cline{4-6}\cline{8-10}\cline{12-14}\cline{16-18}
& & &  \scriptsize Tot & \scriptsize LR & \scriptsize SR &&  \scriptsize Tot &\scriptsize LR & \scriptsize SR && \scriptsize Tot & \scriptsize  LR &\scriptsize SR &&\scriptsize Tot &\scriptsize LR & \scriptsize SR &\\
\hline
\scriptsize$A$&\scriptsize($zz$)&& \scriptsize +0.0008 &\scriptsize +0.0340 &\scriptsize $-$0.0331& &\scriptsize +0.0003 &\scriptsize  +0.0319 &\scriptsize  $-$0.0316 &&\scriptsize $-$0.0020 & \scriptsize +0.0219 & \scriptsize $-$0.0239 &&\scriptsize +0.0211 &\scriptsize +0.0125 &\scriptsize +0.0085 \\

\hline
\scriptsize$B$&\scriptsize($zz$)&&\scriptsize +0.3161 &\scriptsize $-$0.4004 &\scriptsize +0.7165 & &\scriptsize +0.2794 &\scriptsize $-$0.4150 &\scriptsize +0.6945 &&\scriptsize +0.1916 &\scriptsize $-$0.4367&\scriptsize +0.6283 &&\scriptsize +0.1834 &\scriptsize $-$0.5409 &\scriptsize +0.7243 \\

\hline
\scriptsize$O$&\scriptsize($zz$)&&\scriptsize +0.1273 &\scriptsize $-$0.1054 &\scriptsize +0.2328 & &\scriptsize  +0.1111 &\scriptsize $-$0.1095 &\scriptsize +0.2207 &&\scriptsize +0.0985 &\scriptsize $-$0.1170 &\scriptsize +0.2156 &&\scriptsize +0.0812 &\scriptsize $-$0.1640 &\scriptsize +0.2452 \\
\hline\hline 
\end{tabular}
\caption{ Interatomic force constants (Hartree/bohr$^2$) between different pairs of atoms in their local (||) and Cartesian ($zz$) coordinates. The two different long-range (LR) and short-range (SR) contributions to the IFCs are also reported.}
\label{tab:7}
\begin{tabular}{lcccccccccccccccccccccccccccccccccccccccccccc}
\hline
\hline
& & \multicolumn{3}{c}{\scriptsize LiTaO$_3$}&&\multicolumn{3}{c}{\scriptsize LiNbO$_3$}&&\multicolumn{3}{c}{\scriptsize LiVO$_3$}&&\multicolumn{3}{c}{\scriptsize NaVO$_3$} \\
\cline{3-5}\cline{7-9}\cline{11-13}\cline{15-17}
\scriptsize Atoms && \scriptsize Tot &  \scriptsize LR &  \scriptsize SR &&  \scriptsize Tot & \scriptsize LR & \scriptsize SR &&  \scriptsize Tot &  \scriptsize  LR & \scriptsize SR && \scriptsize Tot &  \scriptsize LR & \scriptsize SR\\
\hline

\scriptsize $A_0-O'_{3}$&\scriptsize ($\parallel$) &\scriptsize $-$0.0150 &\scriptsize +0.0144 &\scriptsize $-$0.0295 && \scriptsize $-$0.0150 &\scriptsize +0.0124 &\scriptsize $-$0.0274&&\scriptsize $-$0.0161 &\scriptsize +0.0068 &\scriptsize $-$0.0229&&\scriptsize $-$0.0077 &\scriptsize +0.0034 &\scriptsize $-$0.0111\\
&\scriptsize ($zz$) &\scriptsize  $-$0.0049 &\scriptsize $-$0.0116 &\scriptsize +0.0066 &&\scriptsize $-$0.0047 &\scriptsize $-$0.0103 &\scriptsize +0.0056 &&\scriptsize $-$0.0030 &\scriptsize $-$0.0065 &\scriptsize +0.0034&&\scriptsize $-$0.0022 &\scriptsize $-$0.0037 &\scriptsize +0.0015 \\

&\scriptsize dist. &\scriptsize (1.97)& & & &\scriptsize (1.96) &  &  &&\scriptsize (1.96) & &&&\scriptsize (2.39) &&&\\

\hline

\scriptsize $A_0-O_{1}$& ($\parallel$)& \scriptsize  +0.0057 & \scriptsize +0.0044 & \scriptsize +0.0013&& \scriptsize +0.0055 &\scriptsize +0.0037 & \scriptsize +0.0017&& \scriptsize +0.0047&\scriptsize  +0.0027  &\scriptsize  +0.0020&&\scriptsize +0.0007 &\scriptsize +0.0025 &\scriptsize $-$0.0018\\
& \scriptsize ($zz$)  & \scriptsize +0.0012 & \scriptsize +0.0001 & \scriptsize +0.0011&& \scriptsize +0.0013 &\scriptsize $-$0.0002 & \scriptsize +0.0014&&  \scriptsize +0.0013  &\scriptsize $-$0.0003 &\scriptsize +0.0016&& \scriptsize $-$0.0011 &\scriptsize +0.0001 &\scriptsize $-$0.0013\\

 &\scriptsize dist.&\scriptsize (2.79)& & & &\scriptsize (2.79) && & & \scriptsize(2.61) &&&&\scriptsize(2.62)&& \\

\hline
\scriptsize $B_0-O_{1}$ &\scriptsize ($\parallel$) &\scriptsize $-$0.0573  &\scriptsize +0.2994 &\scriptsize $-$0.3568&&\scriptsize $-$0.0418 &\scriptsize +0.3021 &\scriptsize $-$0.3439&&\scriptsize +0.0119 &\scriptsize +0.3136 &\scriptsize $-$0.3017&&\scriptsize  +0.0334 &\scriptsize  +0.3709 &\scriptsize  $-$0.3379\\
&\scriptsize  ($zz$) &\scriptsize $-$0.0250 &\scriptsize +0.0726 &\scriptsize $-$0.0976&& \scriptsize $-$0.0174 &\scriptsize +0.0750 &\scriptsize $-$0.0924&& \scriptsize +0.0052 & \scriptsize +0.0779 &\scriptsize $-$0.0727&& \scriptsize  +0.0105 &\scriptsize  +0.0969 &\scriptsize $-$0.0865\\

&\scriptsize dist. &\scriptsize (1.97)&&&&\scriptsize(1.98) &&&&\scriptsize(1.85) &&&&\scriptsize (1.85)&&\\

\hline

\scriptsize$B_0-O'_{3}$&\scriptsize ($\parallel$)&\scriptsize +0.0189 &\scriptsize +0.0189  &\scriptsize +0.0000 &&\scriptsize +0.0182 & \scriptsize +0.0182 &\scriptsize +0.0000&&\scriptsize +0.0162  &\scriptsize +0.0162 &\scriptsize +0.0000&&\scriptsize +0.0149 &\scriptsize +0.0149 &\scriptsize +0.0000 \\ 
&\scriptsize ($zz$) &\scriptsize $-$0.0155 &\scriptsize $-$0.0155 &\scriptsize +0.0000&& \scriptsize $-$0.0156 &\scriptsize $-$0.0156 &\scriptsize +0.0000&& \scriptsize $-$0.0144 & \scriptsize $-$0.0144 &\scriptsize +0.0000&& \scriptsize $-$0.0117&\scriptsize $-$0.0117 &\scriptsize +0.0000\\
        
&\scriptsize dist&\scriptsize (3.75) &&&&\scriptsize (3.75) &&&&\scriptsize (3.63) &&&&\scriptsize(3.99)&&\\

\hline
\scriptsize $A_0-A'_0$&\scriptsize ($\parallel$) &\scriptsize $-$0.0013  &\scriptsize $-$0.0013 & \scriptsize +0.0000&& \scriptsize $-$0.0011  &\scriptsize $-$0.0011 &\scriptsize $-$0.0000&&\scriptsize $-$0.0007  &\scriptsize  $-$0.0006 &\scriptsize $-$0.0001&&\scriptsize  $-$0.0007  &\scriptsize $-$0.0004 &\scriptsize $-$0.0003\\
&\scriptsize ($zz$) &\scriptsize $-$0.0000 &\scriptsize $-$0.0000 &\scriptsize $-$0.0000&&\scriptsize $-$0.0000 &\scriptsize $-$0.0000 &\scriptsize $-$0.0000&& \scriptsize $-$0.0002 & \scriptsize +0.0000 &\scriptsize $-$0.0002&&\scriptsize $-$0.0003 &\scriptsize $-$0.0000 &\scriptsize $-$0.0003\\

&\scriptsize dist &\scriptsize (3.75) &&&&\scriptsize (3.76) &&&&\scriptsize (3.56) &&&&\scriptsize (3.69) &&\\

\hline

\scriptsize$A_0-A_1$& \scriptsize ($\parallel$) & \scriptsize $-$0.0002 &\scriptsize $-$0.0002  &\scriptsize $-$0.0000&&\scriptsize $-$0.0002 &\scriptsize $-$0.0002 &\scriptsize +0.0000&&\scriptsize $-$0.0001 &\scriptsize $-$0.0001  &\scriptsize +0.0000  
&&\scriptsize $-$0.0000 &\scriptsize $-$0.0000 &\scriptsize +0.0000\\
&\scriptsize ($zz$) &\scriptsize $-$0.0002 &\scriptsize$-$0.0002 &\scriptsize +0.0000&& \scriptsize $-$0.0002 &\scriptsize $-$0.0002 &\scriptsize +0.0000&& \scriptsize $-$0.0001  &\scriptsize $-$0.0001 &\scriptsize +0.0000
&& \scriptsize $-$0.0000  &\scriptsize $-$0.0000 &\scriptsize +0.0000\\

& \scriptsize dist. &\scriptsize(6.82)&&&&\scriptsize (6.82) &&&&\scriptsize (6.35) &&&&\scriptsize (6.39) &&\\
  
\hline

\scriptsize $B_0-B'_0$&\scriptsize ($\parallel$)&\scriptsize $-$0.0836 &\scriptsize $-$0.0666 &\scriptsize $-$0.0169&&\scriptsize $-$0.0847&\scriptsize $-$0.0668 & \scriptsize $-$0.0178&&\scriptsize $-$0.0925 &\scriptsize $-$0.0663 &\scriptsize  $-$0.0261&&\scriptsize $-$0.0991 &\scriptsize $-$0.0658 &\scriptsize $-$0.0333\\
&\scriptsize ($zz$) & \scriptsize $-$0.0243 &\scriptsize $-$0.0014 &\scriptsize $-$0.0228&& \scriptsize $-$0.0253 &\scriptsize $-$0.0009 &\scriptsize $-$0.0243&& \scriptsize  $-$0.0326 &\scriptsize  +0.0002 &\scriptsize  $-$0.0328&& \scriptsize$-$0.0353 & \scriptsize $-$0.0016 &\scriptsize $-$0.0336\\

&\scriptsize dist.&\scriptsize (3.75)&&&&\scriptsize(3.76) &&&&\scriptsize (3.56) && &&\scriptsize (3.69) &&\\

\hline
\scriptsize $B_0-B_1$ & ($\parallel$)& \scriptsize $-$0.0125  &\scriptsize $-$0.0125  &\scriptsize +0.0000&&\scriptsize $-$0.0129 &\scriptsize $-$0.0129 &\scriptsize +0.0000&&\scriptsize $-$0.0135   &\scriptsize  $-$ 0.0135  &\scriptsize +0.0000&&\scriptsize $-$0.0118 &\scriptsize $-$0.0118 &\scriptsize 0.0000\\
&\scriptsize ($zz$) & \scriptsize $-$0.0125  &\scriptsize $-$0.0125  &\scriptsize +0.0000&& \scriptsize $-$0.0127 & \scriptsize $-$0.0129 & \scriptsize +0.0000&& \scriptsize $-$0.0135  &  \scriptsize $-$0.0135   &\scriptsize +0.0000&& \scriptsize $-$0.0118 & \scriptsize $-$0.0118 &\scriptsize 0.0000\\

&\scriptsize dist.&\scriptsize (6.82) & &&&\scriptsize(6.82) &&&&\scriptsize(6.35) &&&&\scriptsize (6.39) &&\\

\hline

\scriptsize $A_0-B_{0}$&\scriptsize ($\parallel$) &\scriptsize $-$0.0134 &\scriptsize $-$0.00134 &\scriptsize +0.0000&& \scriptsize $-$0.0124 &\scriptsize $-$0.0124 &\scriptsize +0.0000&&\scriptsize $-$0.0097 &\scriptsize $-$0.0097 &\scriptsize 0.0000&&\scriptsize $-$0.0078 &\scriptsize $-$0.0078 &\scriptsize 0.0000\\
&\scriptsize ($zz$) &\scriptsize $-$0.0134 &\scriptsize $-$0.0134 &\scriptsize +0.0000&& \scriptsize $-$0.0124 &\scriptsize $-$0.0124 &\scriptsize +0.0000&&\scriptsize $-$0.0097 &\scriptsize $-$0.0097 &\scriptsize +0.0000&&\scriptsize $-$0.0078 &\scriptsize $-$0.0078 &\scriptsize +0.0000\\

&\scriptsize dist. &\scriptsize (3.41)& & & &\scriptsize (3.41) &  &  &&\scriptsize (3.17 int20) & &&&\scriptsize (3.19) &&&\\

\hline
\scriptsize $O_{1}-O_{4}$ &\scriptsize ($\parallel$)& \scriptsize $-$0.0311 &\scriptsize $-$0.0441 &\scriptsize +0.0079&&\scriptsize $-$0.0334 &\scriptsize $-$0.0418 &\scriptsize +0.0084&&\scriptsize $-$0.0451 &\scriptsize $-$0.0425 &\scriptsize $-$0.0026&&\scriptsize $-$0.0543 &\scriptsize $-$0.0427 &\scriptsize $-$0.0116\\
&\scriptsize ($zz$) & \scriptsize $-$0.0212 &\scriptsize $-$0.0266 &\scriptsize +0.0053&& \scriptsize $-$0.0217 &\scriptsize $-$0.0273 &\scriptsize +0.0056&& \scriptsize $-$0.0301 &\scriptsize $-$0.0277 &\scriptsize $-$0.0024&&\scriptsize$-$0.0347 &\scriptsize $-$0.0277 &\scriptsize $-$0.0070\\

&\scriptsize dist.&\scriptsize(2.79) &&&&\scriptsize(2.79) &&&&\scriptsize(2.61) &&&&\scriptsize(2.62) &&\\
 
\hline  
\hline
\end{tabular}
\end{table*}

Several unstable transverse optic ($TO$) phonon modes are revealed at $\Gamma$ (Table~\ref{tab:2}), in agreement with previous DFT calculations~\cite{veithen2002, parlinski, toyoura, Pengfei2016}. For LiTaO$_3$, there is one polar mode $\Gamma_2^{-}$ (A$_{2u}$) and one  antipolar mode $\Gamma_2^{+}$ (A$_{2g}$) at higher frequency. For LiNbO$_3$, in addition to $\Gamma_2^{-}$ and $\Gamma_2^{+}$ modes, there is  $\Gamma_3^{-}$ (E$_{u}$) polar mode at higher frequency, doubly degenerated in the $xy$-direction. For LiVO$_3$, there are two  $\Gamma_2^{-}$ modes, one $\Gamma_2^{+}$ and one doubly degenerate $\Gamma_3^{-}$ mode. For NaVO$_3$ there is one $\Gamma_2^{-}$ mode, one $\Gamma_3^{-}$ mode and two antipolar $\Gamma_3^{+}$ (E$_{g}$) doubly degenerated in the $xy$-direction. It is worth noting that in LiVO$_3$ and NaVO$_3$, the modes are highly unstable compared to LiTaO$_3$ and LiNbO$_3$. 
Beside these unstable $TO$ modes, we also found one unstable polar $LO1$ mode of $\Gamma_2^{-}$ symmetry for LiTaO$_3$, LiNbO$_3$ and LiVO$_3$ ($\omega_{LO1}$ = 28$i$ $cm^{-1}$, 77$i$ and 138$i$, respectively). Such $LO1$ mode is highly stable in NaVO$_3$ ($\omega_{LO1} = 154~cm^{-1}$). \\

In Table~\ref{tab:4}, we reported the eigendisplacements along the $z$-direction of the $\Gamma_2^{-}$-$TO$ modes (labeled $TO1$, $TO2$ and $TO3$) and $LO1$ mode, together with their associated frequencies. Mode effective charges $\bar{Z}^{*}_m$ are also reported: the $\alpha$-component of the mode effective charge vector is defined as $\bar{Z}^{*}_{m,\alpha}=\frac{\sum_{k\beta}Z^*_{k,\alpha\beta}\eta_{m\bm q=0}(k\beta)}{[\sum_{k\beta}\eta^*_{m\bm q=0}(k\beta)\eta_{m\bm q=0}(k\beta)]^{1/2}}$~\cite{Gonze1997}, where $\eta_{m\bm q=0}$ is the eigendisplacement associated to the mode $m$ at the $\Gamma$-point and $Z^*_{k,\alpha\beta}$ are the Born effective charges -or transverse charges $\bm{Z}^{*(T)}$- for the $TO$-modes, and the Callen effective charges -or longitudinal charges $\bm{Z}^{*(L)}$- for the $LO$-modes; $\bm{Z}^{*(L)}$ is directly related to the $\bm{Z}^{*(T)}$ via the electronic dielectric tensor $\bm \epsilon_{\infty}$, i.e.  $\bm{Z}_k^{*(L)}=\bm\epsilon^{-1}_{\infty}\bm{Z}_k^{*(T)}$~\cite{Ghosez1998, callen}. Complete $\bm{Z}^{*(T)}$ tensor and $\eta_{m\bm q=0}$ components are reported in Supplemental Material (See~\cite{supplemental}).

The eigendisplacements associated to the unstable $TO1$ modes showed that A- and B-sites cations displace in phase along the trigonal axis, but in antiphase with respect to the oxygens.
These modes exhibit a large mode effective charge, mostly resulting from the anomalous Born effective charges on the B atoms and oxygens (A atoms show values close to their nominal ionic charge), as reported in Table~\ref{tab:3}. In particular, a very large displacement of Li atoms characterizes $TO1$ in LiTaO$_3$ and LiNbO$_3$, in contrast to LiVO$_3$ and NaVO$_3$, where it is the V atoms at the B-site that move the most.
In these latter, the dominant B-site motion in $TO1$,
combined with the very anomalous BEC on V and O atoms, produce the extremely large $\bar{Z}^{*}_{TO1}$ observed in LiVO$_3$ and NaVO$_3$ compared to LiTaO$_3$ and LiNbO$_3$.
The eigendisplacements associated to the $TO2$ mode, that is unstable only in LiVO$_3$, showed a large motion of A atoms in the four systems. However, the A-cations displace in-phase with oxygens in LiTaO$_3$ and LiNbO$_3$ and in anti-phase in LiVO$_3$ and NaVO$_3$; at the opposite, the B-cations displace in anti-phase with oxygens in LiTaO$_3$ and LiNbO$_3$, producing a still large $\bar{Z}^{*}_{TO2}$, while, they displace in-phase with oxygens in LiVO$_3$ and NaVO$_3$, causing the vanishing $\bar{Z}^{*}_{TO2}$. $TO3$ modes are highly stable in all the four systems. 

The eigendisplacements associated to the unstable $LO1$ mode showed 
a dominant A atoms motion in anti-phase with both B atoms and oxygens (A-O motion seems in-phase in LiTaO$_3$, but O contribution is quasi negligible). Such A-site driven character, already suggest the active role of the Li-cation in driving the $LO$ instability. 

In order to estimate the correlation between $LO1$ and the $TO$ modes, we calculated the overlap matrix elements between the corresponding eigendisplacements as $\langle \eta^{LO1}|M|\eta^{TO}\rangle$ (projection of $LO1$ mode on the basis of the $TO$ modes, as in ~\cite{djani12}) where $M=M_k\delta_{kk'}$ with $M_k$ the mass of atom $k$ . 
The results, reported in Table~\ref{tab:5}, show that $LO1$ eigendisplacements are mainly associated to the $TO$ modes displaying dominant A atoms motions. In particular, in LiTaO$_3$ and LiNbO$_3$, $LO1$ results from a mixing between $TO1$ and $TO2$, while it is mainly associated to $TO2$ in LiVO$_3$ and NaVO$_3$. It is thus important to emphasize that there is no one-to-one correspondence between $LO1$ and $TO1$ as considered by Li et $al$ in Ref.~\cite{Pengfei2016}. 

Moreover, in the aim of quantifying the balance between the dipole-dipole long range interactions and the short-range interactions behind the above discussed unstable modes, we  decomposed the phonons frequencies into  two contributions, $ie$. $\omega^2$= $\omega_{LR}^2$ + $\omega_{SR}^2$, as discussed in Ref.~~\cite{Ghosez_1996}.
From this decomposition reported in Table~\ref{tab:4}, it is interesting to note that only $TO1$ instability in LiTaO$_3$ and $TO2$ one in LiVO$_3$ originate from a global destabilizing SR interactions ($\omega_{SR}^2 < 0$), while, all the other instabilities originate from destabilizing LR interactions ($\omega_{LR}^2 < 0$). Nevertheless, the fact that LiNbO$_3$ also hosts the $LO$ instability without showing global unstable SR interactions, suggests that this is not the necessary condition for hyperFE, but rather, the specific destabilizing SR  interatomic interactions associated to the Li motion (dominant Li motion in $LO1$ mode is a common feature in LiTaO$_3$, LiNbO$_3$ and LiVO$_3$, see above).

Accordingly, to shed light on the necessary conditions that make the A-site motion active in the destabilization of the $LO$ mode, we examined distinct atomic interactions, through the analysis of the on-site and interatomic force constants, calculated in the $R\bar{3}c$ paraelectric phase. Interestingly,  we found that the ($zz$) component (out-of-plane direction) of the on-site force constant for Li is vanishingly small in LiTaO$_3$ and LiNbO$_3$ and turns out to be negative for LiVO$_3$ (see Table~\ref{tab:6}). In particular, we noticed a negative contribution, \emph{i.e.} destabilizing, of the SR part for the three systems displaying the unstable $LO$ mode. The other components of the Li on-site force constants and all those of Na, B and O are large and positive.

The SR nature of the $LO1$ instability and the active role played by Li are also  highlighted from the examination of the interatomic force constants reported in Table~\ref{tab:7}. The $A_0$-O$_1$ IFC (equivalent to the $A_0$-O$_{2,3}$), related to the interaction between the under-coordinated A atom and the out-of-plane oxygens toward which it tends to displace, is destabilizing with respect to both the LR and the SR forces (positive values) in LiTaO$_3$, LiNbO$_3$ and LiVO$_3$ systems. On one hand, the destabilizing LR dipole-dipole interaction contributes to the instability of the $TO1$ mode; on the other hand the destabilizing SR interaction is responsible for the $LO1$ instability and its associated eigendisplacement dominated by Li motion. The destabilizing  $A_0$-O$_1$ interaction confirms also the correlated Li-O motion reported in Ref~\cite{inbar}.  Noteworthy, the SR part of the $A_0$-O'$_3$ IFC, related to the interaction between A atom and in-plane oxygens, is also destabilizing along the $z$-direction, but not strong enough to overcome the stable LR part. In NaVO$_3$, in which the $LO1$ is stable, the scenario is indeed different: 
it is only the LR dipole dipole interaction that is destabilizing and thus driving the cooperative Na-O polar motion (SR part of $A_0$-O$_1$ and $A_0$-O'$_3$ IFCs is negative). 

The interatomic force constants between the B atoms and the oxygens exhibit a destabilizing LR dipole-dipole interaction as standard ferroelectric perovskites~\cite{PhysRevB.60.836,PhysRevB.97.174108}; the LR forces are destabilizing along the direction parallel to the $B-O$ bonds. In particular, LiVO$_3$ and NaVO$_3$ exhibit much more unstable $TO1$ mode than LiTaO$_3$, LiNbO$_3$ that is characterized by dominant anti-phase displacement of V and O atoms and giant mode effective charges (due to anomalous BEC on V and O atoms as discussed in previous paragraph).


 We finally noted negative interatomic force constants between near and next-near neighboring A and B atoms, meaning that the motion of A and/or B atoms along the $A-B$ chain would be in-phase, propagating thus the polar distortions along this chain ($i.e$. along the trigonal axis)~\cite{Ghosez1998}. 

\subsection{The effect of isotropic pressure}

\begin{figure*}[t]
\centering
\includegraphics[width=12cm]{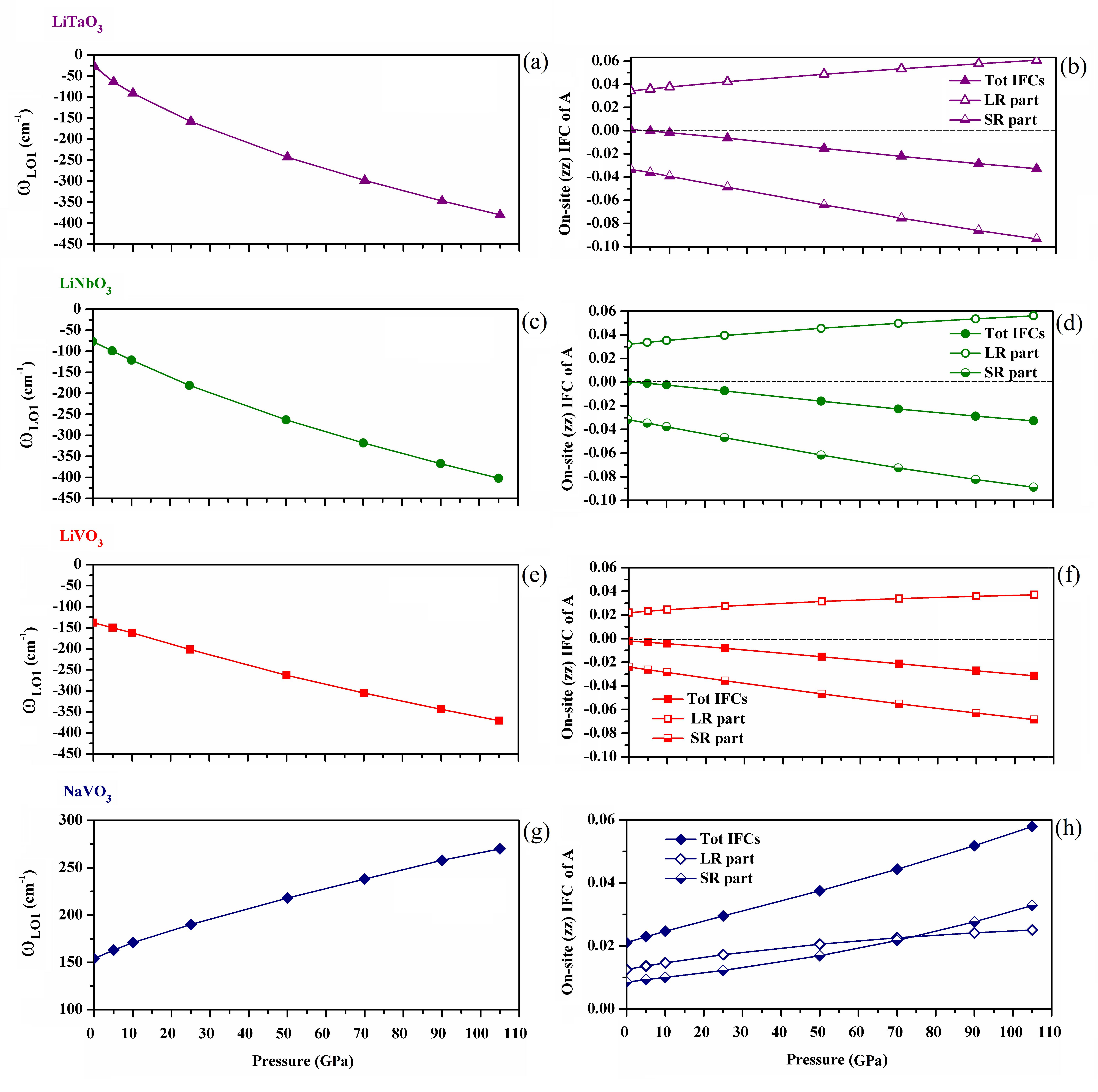}
\caption{(a,c,e,g) Evolution of $LO1$ mode frequency with pressure. (b,d,f,h) Evolution of A atoms total ($zz$) on-site IFC with pressure, together with its LR and SR parts. 
}
\label{Fig3}
\centering
\includegraphics[width=\textwidth]{Fig4-new.pdf}
\caption{(a,b,c) Evolution of Li displacement in $TO1$, $TO2$ and $TO3$ polar modes under increasing pressure. (d,e,f) Evolution of the overlap the $LO1$ mode with $TO1$, $TO2$ and $TO3$ modes with pressure. } 
\label{Fig4}
\end{figure*}

Ferroelectricity is known to be highly sensitive to external pressure; in particular,  it was highlighted that, in the high-pressure ferroelectricity, a crucial role is played by SR range interactions, which become destabilizing~\cite{samara, Kornev2005,Bousquet2006}.
Based on that, it appeared necessary to explore the effect of  isotropic pressure (compressive strain) on hyperferroelectricity, which is, as showed in previous section, mainly driven by destabilizing SR forces on A atoms.



Interestingly, the $LO1$ instability increases as a function of pressure in LiTaO$_3$, LiNbO$_3$ and LiVO$_3$ [Fig.~\ref{Fig3}(a,c,e)]; opposite trend is observed in NaVO$_3$ [Fig.~\ref{Fig3}(g)]. The destabilizing SR range forces acting on Li atoms are in fact strengthened by the compressive strain, as shown in Fig.~\ref{Fig3}(b,d,f); in particular, we observe that the on-site force constants of Li atoms become more and more negative with increasing pressure, following the evolution of its SR part. At opposite, pressure increases the stiffness of Na atoms in NaVO$_3$: the on-site force constant associated to Na becomes more and more positive, as shown in Fig.~\ref{Fig3}(h).

Noteworthy, the contribution of the $TO1$ mode to the $LO1$ mode in LiTaO$_3$, LiNbO$_3$ and LiVO$_3$ also increases with pressure, as clearly shown from the evolution of the overlap matrix illustrated in Fig.~\ref{Fig4}(d,e,f). 
This is correlated to the increasing Li-displacement in the $TO1$-eigendisplacement (see Fig.~\ref{Fig4}(a,b,c)) and is consistent with the fact that $LO1$ is associated to the $TO$ modes exhibiting a large motion of the frustrated atom, as discussed in the previous section. 

More interestingly, we observed that the $LO1$ mode effective charges also increase with pressure for the three Li-based compounds (see Table~\ref{tab:8}); this is due to the increasing, anti-phase oxygens motion in the $LO1$ eigendisplacement. Since the mode effective charges are giving, by construction, an idea about the polarity of the mode, 
this result suggests that an external pressure could enhance the hyperferroelectric polarization. The polarization under open circuit conditions ($\mathscr{D}=0$), not calculated in this work, is in fact found to be very small at 0GPa (see ref.~\cite{Garrity2014, Pengfei2016, halides}).  


\begin{table}[t]
\begin{center}
\squeezetable
\caption{Effect of Increasing pressure on $LO1$ mode frequency, on $LO1$  eigenvectors along $z$-direction and on longitudinal mode effective charges ($\bar{Z}^{*L}_{LO1}$).}
\label{tab:8}
\begin{tabular}{lcccccccccccccccccccc}
\hline
\hline
P(Gpa)  & &  $\omega_{LO1}$ & A & B & O$_{1/2/3}$ && $\bar{Z}^{*L}_{LO1}$ \\
\hline
LiTaO$_3$ & & & & & & & & & & &\\
0 & & [28$i$]& +0.2594 & $-$0.0105 &+0.0021 && 0.077 & \\
5 & & [64$i$]& +0.2597 & $-$0.0101 & +0.0006&& 0.093 & \\
10 & & [91$i$]& +0.2598 & $-$0.0096 & $-$0.0010 && 0.110 & \\
25 & & [158$i$] & +0.2591 & $-$0.0083 &$-$0.0061 && 0.156 &\\
50 & & [243$i$] & +0.2549 & $-$0.0059 & $-$0.0145 && 0.224 & \\
70 & & [299$i$] & +0.2491 & $-$0.0040 & $-$0.0207&& 0.270 &\\
90 & & [347$i$]& +0.2415 & $-$0.0022 & $-$0.0265 && 0.310 &\\
105 & & [381$i$]& +0.2347 & $-$0.0008 & $-$0.0306 && 0.336&\\
\hline
LiNbO$_3$ & & & & & & & & & & &\\
0 & & [77$i$]& +0.2570& $-$0.0162 & $-$0.0057 && 0.053 &\\
5 & & [99$i$]& +0.2571 & $-$0.0155 & $-$0.0072&& 0.068 & \\
10 & & [121$i$]& +0.2570 & $-$0.0147 & $-$0.0087 && 0.084 &\\
25 & & [181$i$] & +0.2556 & $-$0.0123 &$-$0.0132 && 0.127 &\\
50 & & [263$i$] & +0.2501 & $-$0.0081& $-$0.0205 && 0.192 &\\
70 & & [318$i$]& +0.2429 & $-$0.0046 & $-$0.0262 && 0.236 & \\
90 & & [368$i$]& +0.2334 &$-$0.0011 &$-$0.0315 && 0.274&\\
105 & & [402$i$]& +0.2248 & +0.0014 &$-$0.0353 && 0.299&\\
\hline
LiVO$_3$ & & & & & & & & & & & \\
0 & & [138$i$]& +0.2559& $-$0.0258 & $-$0.0096 && -0.008 &\\
5 & & [150$i$]& +0.2562& $-$0.0250 & $-$0.0105&& 0.001 & \\
10 & & [162$i$]& +0.2565& $-$0.0243 & $-$0.0113 && 0.010 &\\
25 & & [202$i$] & +0.2570 & $-$0.0221 & $-$0.0136 && 0.033 &\\
50 & & [263$i$] &+0.2569 &$-$0.0188 &  $-$0.0172 && 0.064 &\\
70 & & [305$i$]& +0.2561 & $-$0.0162 &$-$0.0198 && 0.085 &\\
90 & & [344$i$]& +0.2549 & $-$0.0137 & $-$0.0223 && 0.104 &\\
105 & & [371$i$]& +0.2536 &$-$0.0119& $-$0.0241 && 0.116 &\\
\hline\hline 
\end{tabular}
\end{center}
\end{table}

\begin{table*}[htpb!]
\begin{center}
\squeezetable
\caption{On-site and interatomic force constants in (Hartree/bohr$^{2}$) related to different atoms of the hexagonal ABC systems calculated in the $P6_3/mmc$ (194) paraelectric phase.}
\label{tab:9}
\begin{tabular}{lccccccccccccccccccccccc}
\hline
\hline
        & & & \multicolumn{3}{c}{\scriptsize LiBeSb \scriptsize(LO=16$i$, TO$_1$=126$i$)}&&\multicolumn{3}{c}{\scriptsize LiZnP\scriptsize(LO=82, TO$_1$=82$i$)}&&\multicolumn{3}{c}{\scriptsize LiZnAs \scriptsize(LO=36$i$, TO$_1$=83$i$)}&&\multicolumn{3}{c}{\scriptsize NaMgP \scriptsize(LO=155, TO$_1$=113$i$)} \rule[-1ex]{0pt}{3.5ex} \\
\cline{4-6}\cline{8-10}\cline{12-14}\cline{16-18}
& & &  \scriptsize Tot & \scriptsize LR & \scriptsize SR &&  \scriptsize Tot &\scriptsize LR & \scriptsize SR && \scriptsize Tot & \scriptsize  LR &\scriptsize SR &&\scriptsize Tot &\scriptsize LR & \scriptsize SR &\\
\hline
\scriptsize $A$&\scriptsize($zz$)&& \scriptsize +0.0178 &\scriptsize  +0.0006 &\scriptsize +0.0172 & &\scriptsize +0.0172 &\scriptsize $-$0.0010 &\scriptsize +0.0182 &&\scriptsize  +0.0149  & \scriptsize $-$0.0011  & \scriptsize  +0.0160 &&\scriptsize +0.0297  &\scriptsize +0.0013 &\scriptsize +0.0284 \\

\hline
\scriptsize $B$ &\scriptsize($zz$)&&\scriptsize $-$0.0011 &\scriptsize  +0.0045 &\scriptsize $-$0.0056  & &\scriptsize +0.0022 &\scriptsize +0.0100 &\scriptsize $-$0.0078 &&\scriptsize $-$0.0042  &\scriptsize  +0.0082 &\scriptsize $-$0.0124 &&\scriptsize +0.0023 &\scriptsize +0.0066  &\scriptsize $-$0.0043  \\

\hline
\scriptsize $C$ &\scriptsize($zz$)&&\scriptsize +0.0158 &\scriptsize +0.0047 &\scriptsize +0.0111 & &\scriptsize +0.0061 &\scriptsize +0.0059 &\scriptsize +0.0001 &&\scriptsize +0.0009 &\scriptsize +0.0043 &\scriptsize $-$0.0034 &&\scriptsize +0.0103 &\scriptsize +0.0087 &\scriptsize +0.0016 \\

\hline

\scriptsize $B-C_{0}$& \scriptsize($zz$)&&  \scriptsize +0.0003 & \scriptsize $-$0.0025 & \scriptsize +0.0029 && \scriptsize +0.0004  &\scriptsize $-$0.0040  & \scriptsize +0.0045 &&  \scriptsize  +0.0022  &\scriptsize $-$0.0032 &\scriptsize +0.0054  && \scriptsize $-$0.0009   &\scriptsize $-$0.0050  &\scriptsize +0.0041 \\

&\scriptsize dist.&&\scriptsize (2.36)& & & &\scriptsize (2.31) && & & \scriptsize(2.40) &&&&\scriptsize(2.53)&& \\
\hline

\scriptsize $B-C_{1}$& \scriptsize($zz$)&&  \scriptsize +0.0019 & \scriptsize  +0.0018 & \scriptsize +0.0001 && \scriptsize +0.0016  &\scriptsize  +0.0018 & \scriptsize $-$0.0002 &&  \scriptsize +0.0013  &\scriptsize +0.0013 &\scriptsize $-$0.0000 && \scriptsize +0.0046  &\scriptsize +0.0048  &\scriptsize $-$0.0002 \\

 &\scriptsize dist.&&\scriptsize (3.87)& & & &\scriptsize (3.64) && & & \scriptsize(3.73) &&&&\scriptsize(3.63)&& \\          
\hline\hline 
\end{tabular}
\end{center}
\end{table*}

\section{Discussion} 

LiTaO$_3$, LiNbO$_3$, LiVO$_3$ and NaVO$_3$ compounds all exhibit polar instabilities; in particular, the unstable $TO1$ modes in LiTaO$_3$, LiNbO$_3$ and the $TO2$ one in LiVO$_3$ are characterized by dominant anti-phase Li-O displacements (Table~\ref{tab:4}). The A-cation in these ABO$_3$-LiNbO$_3$-type systems, experiences in fact an electrostatic frustration due to its under-coordination in the $R\bar{3}c$ paraelectric phase~\cite{xiang2014}: the off-centering of Li from its central position in the O-triangle toward the three out-of-plane oxygens (Fig. 2) optimizes Li-coordination from III, in the paraelectric $R\bar{3}c$ phase, to VI in the ferroelectric $R3c$ phase. 

The instability of the polar $LO1$ mode is ascribed, in one hand, to this frustration, as the $LO1$ mode is mainly driven by the A-site motion and, on the other hand, to the small size of the frustrated cation. The $LO1$ mode is in fact unstable only in the Li-based compounds, where Li has much smaller size than Na (0.76 vs 1.02 \AA~ respectively, for six-coordinated cations~\cite{Shannon1973}). 
Moreover, by analysing the on-site (Table~\ref{tab:6}) and the interatomic (Table~\ref{tab:7}) force constants, we found out that the $LO1$ instability is driven by short-range interactions: LiTaO$_3$, LiNbO$_3$, LiVO$_3$ exhibit destabilizing Li-O interactions with dominant contribution coming from SR forces; in particular, this leads to vanishingly small or even negative Li on-site force constants, that is not observed in the case of NaVO$_3$.  

\begin{figure}[t]
\includegraphics[width=8.2cm,keepaspectratio=true]{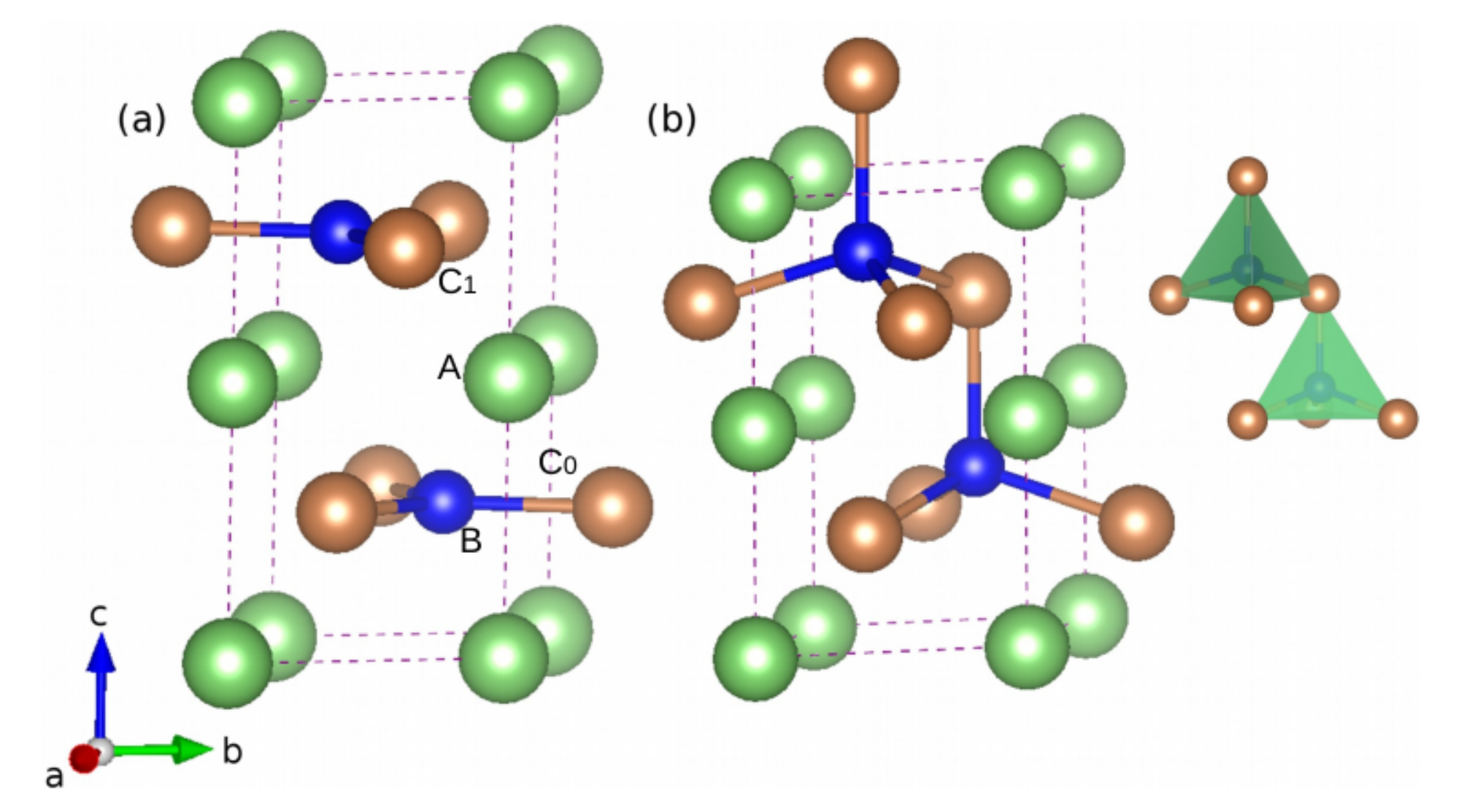}
\caption{ ABC-hexagonal cell in the paraelectric  $P6_3/mmc$ and polar $P6_3mc$(186) phases.}
 \label{ABC_structure}
\end{figure}

Our findings on ABO$_3$-LiNbO$_3$-type systems are also confirmed in the prototype hyperFEs, the ABC-hexagonal systems~\cite{Garrity2014}. In the ABC-hexagonal systems, it is the small B cation, under-coordinated in the paraelectric $P6_3/mmc$ phase, to experience a structural frustration (see Fig.~\ref{ABC_structure}(a)): as for Li in the LiNbO$_3$-type systems, the B cation sits a the center of a triangle formed by three near-neighbor C-anions; its off-centering towards the apical C-anion improves (along the $z$-direction) its coordination from III, in the paraelectric $P6_3/mmc$ phase, to IV, in the ferroelectric $P6_3mc$ phase (see Fig.~\ref{ABC_structure}). By exploring the on-site and B-C interatomic force constants in LiBeSb, LiZnAs, LiZnP and NaMgP, taken as representative examples, we found out that the $LO1$ instability occurs in the compounds which exhibit some dominant destabilizing SR interactions, like LiBeSb and LiZnAs. In these systems in fact the $zz$ component of the on-site force constant of the frustrated B-cation is negative and the B-C$_0$ interaction is positive (Table~\ref{tab:9}), in line with what we argued for the LiNbO$_3$-type systems. 

We also suggested the possible enhancement of hyperferroelectricity in LiNbO$_3$-type compounds by applying an external isotropic pressure (compressive strain). Indeed, pressure strengthens the short-range forces, as observed from the trend of the $zz$- component of the on-site force constants of Li in LiTaO$_3$, LiNbO$_3$, LiVO$_3$, which becomes more and more negative; in turn, this produces further softening of the $LO1$ frequency, which becomes, monotonically, more and more unstable (Fig.~\ref{Fig3}). Interestingly, the Callen mode effective charge associated to the $LO1$ mode increases with pressure (Table~\ref{tab:8}), suggesting thus a possible increase in the hyperferroelectric polarization.

Finally, it is also noteworthy to mention that the current literature is considering that the unstable $LO$ mode arises from a small $LO-TO$ splitting ~\cite{Garrity2014, Pengfei2016}. This notion is however not conclusive since a one-to-one correspondence between $TO$ and $LO$ modes does not always occur, as in the case of LiNbO$_3$-type systems.
The calculation of the overlap matrix (Table~\ref{tab:5}) showed in fact that only in LiVO$_3$ the unstable $LO$ mode totally corresponds to one unstable $TO$ mode, that is the $TO2$; in LiTaO$_3$ and LiNbO$_3$, the unstable $LO1$ mode is correlated to two $TO$ modes instead (the unstable TO1 and the stable TO2). 
Moreover, the overlap matrix also revealed that the $TO$ modes contributing to the $LO$ one are those characterized by large motion of Li atoms, consistently with the Li-driven character of the structural instability.

\section{Conclusions}

In this paper we have investigated the dynamical properties of LiTaO$_3$, LiNbO$_3$, LiVO$_3$ and NaVO$_3$ compounds by means of first-principles calculations revealing microscopic mechanisms of general validity behind hyperferroelectricity; we have provided in fact a confirmation also for the ABC-hexagonal systems. In particular, we have shown that, beyond the small $LO-TO$ splitting claimed in literature, the $LO$ mode instability is driven by destabilizing short-range forces acting on the small sized cations at the A-site of the ABO$_3$-LiNbO$_3$-type systems and at the B-site of the ABC-hexagonal ones, which experience an electrostatic frustration caused by their under-coordination in their respective centrosymmetric paraelectric phases. The signature of such SR-driven $LO$ instability is a vanishingly small or negative on-site force constant associated to the frustrated cation, which reflect its tendency to displace, combined to destabilizing cation-anion interactions; both effects associated to destabilizing SR forces. Moreover, we have also predicted a possible enhancement of hyperferroelectricity upon external isotropic pressure, which can be suitable for technological applications.



\section{Acknowledgements} 
Authors acknowledge Ph. Ghosez from University of Li\`ege for useful discussions. Computational resources are provided from CDTA cloud platform. M.K. and  H.D. acknowledge support from Algerian-WBI bilateral cooperative project.  D.A. is grateful to S. Picozzi (CNR-SPIN) for the provided time to work on this paper.


\bibliography{biblio}

\end{document}